\documentstyle[12pt,fleqn,epsf]{article}
\begin{document}
\title{
\rightline{\small submitted to Physical Review E}
       Conservation laws in coupled multiplicative random arrays lead 
       to $1/f$ noise
 \footnote{Correspondence should be addressed to M.C.T. (e-mail: teich@bu.edu;\\
           URL: http://ece.bu.edu/ECE/faculty/homepages/teich.html/)}
} 
\author{
    Stefan Thurner,$^1$ Markus C. Feurstein,$^1$   
                     and Malvin C.~Teich$^{1,2}$  \\
    $^1$ {\it Department of Electrical and  Computer Engineering,}\\ 
    {\it Boston University, } {\it  Boston, Massachusetts 02215, USA} \\
    $^2$ {\it Departments of Physics, Biomedical Engineering,  and }\\ 
    {\it Cognitive \& Neural Systems, Boston University,}\\
    {\it   Boston,  Massachusetts 02215, USA} \\
  }
\date{}
\maketitle
\begin{abstract}
\noindent

We consider the dynamic evolution of a coupled array of 
$N$ multiplicative random 
variables. The magnitude of each is constrained by a lower bound $w_0$
and their sum is conserved. Analytical calculation shows that the 
simplest case, $N=2$ and $w_0=0$, exhibits a Lorentzian spectrum which
gradually becomes fractal as $w_0$ increases. Simulation results for larger
$N$ reveal fractal spectra for moderate to high values of $w_0$ 
and power-law amplitude fluctuations at all values. 
The results are applied to estimating the fractal exponents for
cochlear-nerve-fiber action-potential sequences with remarkable success,
using only two parameters. 

\end{abstract}

%

\clearpage

\section{Introduction}
Over the past decade it has become apparent that power-law behavior is
ubiquitous in physical and biological phenomena alike \cite{MAN82,FED88,
SCH91,BAS94}.     
In spite of the importance of developing models to describe these 
phenomena, the literature is rather sparse when it comes to 
first-principle approaches that deal with the origins of fractal 
power-law (1/f) phenomena. Fractal point processes in particular 
have received short 
shrift in this connection; this important class of models is
suitable for describing phenomena such as  
trapping times in amorphous semiconductors, neurotransmitter exocytosis, 
ion-channel opening times, and nerve-fiber action-potential occurrences 
\cite{LOWE95,THUR97}. 

In this paper we develop a plausible dynamic multiplicative stochastic 
model, with well-understood underpinnings, that yields fractal
behavior under certain specified conditions. The model comprises 
interchangeable variables and an explicit 
constraint on the allowed values of its individual elements.
The system is further restricted by a global conservation law. We 
find that this simple construct leads to fractal $1/f$ correlations both 
in the time evolution of the individual variables and in the asymptotic 
distribution of all of the variables. We demonstrate that the fractal 
exponents that emerge from the model are controlled solely by 
the constraints on the individual variables and the size of the array
(number of elements).
The model is applied to a biological system that is in fact 
subject to just such constraints and conservation laws. 
Fractal exponents associated with peripheral mammalian 
neural-spike trains \cite{TEI89,LOWE96,LOWE97} that carry auditory 
information to higher centers in the brain are estimated.
The model requires only 
two parameters, both of which can be determined
physiologically: the number of innervated nerve fibers per 
inner hair cell and the minimum 
neurotransmitter flux per afferent nerve fiber. 

We anticipate that the results developed here
will find application in a broad range of problems in the 
physical and biological sciences.

\section{Model}
\subsection{Coupled Array in the Asymptotic Limit}
We consider a system comprising of a set of $N$ elements, each 
denoted $i$ and characterized by a time-dependent real-valued 
variable $w_i(t)$ (we henceforth refer to this as the 
array of variables). The discrete time evolution of this array of 
stochastic variables is prescribed by 
\begin{equation}
\label{proc}
w_i(t+1)=\lambda_i \, w_i(t),
\end{equation}
where $\lambda$ is a random variable drawn from a probability density  
$\pi(\lambda)$ with compact support. This probability density 
depends on neither on $i$ nor on the actual value of $w_i$, and 
$\lambda$ is independently drawn from $\pi(\lambda)$ for each element 
$i$. Each variable therefore characterizes a branching process 
\cite{HARR89}. 

The elements $i$ are now coupled by imposing a normalization condition
on the sum of their values:
\begin{equation}
\label{norm}
\sum_{i=1}^{N} w_i(t) = N \,\, ,
\end{equation}
so that the mean value of $w_i(t)$ is unity. We further impose the
condition that $w$ is bounded from below for  all $i$:
\begin{equation}
w_i(t) > w_0,
\end{equation}
with $w_0 \geq 0$.

In the limit of large times and large $N$, the distribution 
of the variables $w$, denoted $P(w)$, can be determined with the
help of the corresponding master equation. The result turns out to
exhibit power-law behavior \cite{Levy96}
\begin{equation}
\label{resu}
P(w) \sim w^{-1-1/T} =w^{-\alpha_{\rm{asym}}}\,\, ,
\end{equation}
where $T=1-w_0$ and $\alpha_{\rm{asym}} = 1 + 1/T$; interestingly 
it is independent of the nature of $\pi(\lambda)$.


The validity of Eq. \ref{resu} was verified using simple computer 
simulations of Eq. (\ref{proc}). 
We considered $N$ elements (usually 1024) and for $\pi(\lambda)$,
either a uniform distribution $\pi_{\rm{unif}}=\Theta(\lambda-s)
\cdot \Theta(s+1-\lambda)$ or an exponential distribution
$\pi_{\rm{exp}}=e^{c \lambda}\cdot \Theta(\lambda-s) 
\cdot \Theta(s+1-\lambda)$, where $\Theta(x)$ is the Heaviside function, 
and $s$ represents a non-negative shift parameter.
The set of variables $w_i(t)$, initially set equal to unity, was updated
at the time step ($t'=t+1$) in the following way: each of the $N$ 
stochastic variables $w_i(t)$ was multiplied by a random variable 
$\lambda_i$ drawn from $\pi(\lambda)$ (the $\lambda_i$ were different
samples drawn from a single distribution), and were then subjected to
normalization such that their sum was equal to the total number of 
elements in the array, $N$. In this new sequence all $w_i$ values that  
did not obey the restriction $w_i>w_0$ were multiplied by a new 
random variable $\lambda$, and the whole set $i$ was then normalized 
again. This procedure was iterated as long as there were cases for 
which $w_i \leq w_0$. The final set $w_i$ was then considered to 
be the sequence at the time $t'=t+1$. The free parameters of the 
simulations were the lower bound $w_0$, the form of $\pi$, the shift 
parameter $s$, and the array size $N$.

For relatively 
short time sequences ($3\,000$ time steps), a power-law amplitude
histogram $P(w)$ emerges and the expected  
dependence $\alpha(w_0)=1+1/(1-w_0)$ is qualitatively reproduced.
We confirmed that the form of the distribution $\pi(\lambda)$ indeed
has a minimal effect on the outcome. We therefore restricted our
consideration to $\pi=\pi_{\rm{unif}}$ in the remainder of this paper.
Moreover, we found that bias effects associated with the value of $s$ remained
small if $s\approx 1$; we therefore use $s=1$ througout.

The interesting character of this asymptotic result prompts us to 
examine the time evolution of this system, which we proceed to do in the
next section.

\subsection{Dynamical Evolution in the Coupled Array Model}
The dynamical nature of the multiplicative random model,  
constrained by normalization and a minimum value $w_0$ for all variables, 
is elucidated by rewriting Eq. (\ref{proc}) as a set of coupled 
stochastic differential equations,
\begin{eqnarray} 
\label{set}
\dot w_1(t)& =& (\lambda_1-1) w_1(t) \nonumber \\
\dot w_2(t)& =& (\lambda_2-1) w_2(t) \nonumber \\
&.&   \nonumber \\
&.& \nonumber \\
&.& \nonumber \\
\dot w_N(t)& =& (\lambda_N-1) w_N(t) \nonumber \\
\sum_{i=1}^{N} w_i(t)& =& N, \quad {\rm for\,\, all\,\, } t \nonumber \\
w_i(t)&  >&  w_0 , \,\,  {\rm for\,\, all \,\,} t \,\, {\rm and} \,\,i  \, , 
\end{eqnarray}
where all $\lambda_i$ are drawn from a single distribution $\pi(\lambda)$. 
 
In the next subsection we demonstrate that these Langevin equations
can be analytically solved for an array of size $N=2$ (when $w_0 =0$), 
and that the correlations are exponential. However, power-law behavior
emerges as the level of the constraint $w_0$ increases.

\subsubsection{$N=2$, Analytical Solution}
For a system comprising two elements ($i=1,2$), an analytical solution 
can be found for $w_0=0$ by iteratively solving a master equation. 
The set of equations (\ref{set}) can be reduced to a single equation 
that incorporates the normalization condition. The transition 
probability matrix for the system, which provides the  
conditional probability of obtaining the random variable $w(t+1)$, 
when the starting value is $w(t)$ at the previous time step, is given by 
\begin{equation}
W(w(t+1),t+1|w(t),t) = \frac{ \lambda_1 w(t)}{\lambda_1  w(t)+
                \lambda_2 (2-w(t))}
\end{equation}
where $\lambda_{1,2}$ is a random variable drawn from the 
probability distribution $\pi(\lambda)$. 
Equation (6) is readily rewritten in the form 
\begin{equation}
\label{cond}   
W(w(t+1),t+1|w(t),t) =\frac{1}{x},
\end{equation}
where $x$ is a random variable drawn from the distribution 
$f_x=1+f(z)  (2/w(t) - 1)$. The quantity $f(z)$ is the probability 
distribution of the quotient of the two random 
variables, $z=\lambda_2 / \lambda_1$.  
Using basic relations for the products of random variables \cite{
SPR79}, it can be shown that 
\begin{eqnarray}
f(z)&=& \frac{1}{2z^2} \quad {\rm for} \quad z>1 \nonumber \\
    &=& \frac{1}{2} \quad \quad {\rm for} \quad z<1  \quad . \\
\end{eqnarray}   
Finally,setting $y=1/x$, the probability matrix takes
the form    
\begin{equation}
W(w(t+1),t+1 |w(t),t) = \frac{1}{y^2} \cdot f_x(\frac{1}{y}) 
\end{equation} 
where $y$ is a functional of $f(z)$ and $w(t)$. Computed results
are shown in Fig. 1 for uniform probability distribution $\pi(\lambda)$.

Given knowledge of the probability matrix $W$, all of 
the conditional probabilities $P(|)$ can be computed as a 
function of time by using the time-evolution equation
\begin{eqnarray}
P(w(t+2),t+2|w(t),t)= \nonumber \\
\sum_{w(t+1)} W(w(t+2),t+2|w(t+1),t+1) P(w(t+1),t+1|w(t),t) \,\, ,
\end{eqnarray}
which is seen to be an iterative solution of a discrete master equation. 
The conditional probability converges quite nicely to 
a function that is constant in time, as is understood by recognizing
that the matrix $W$ has three degenerate eigenvalues, $E_1=E_2=E_3=1$. 
All of the other eigenvalues are smaller than unity and vanish under 
repeated applications of $W$. Two of the relevant 
corresponding eigenvectors are trivial; the third is the  
asymptotic probability $P(w,t\rightarrow \infty)$.
 
These probabilities permit the correlation functions to be computed by 
carrying out the integral   
\begin{equation}
<w(0)w(t)>= \int dw  dw' \,  P(w,t,w',0) \, w\, w' \,\, , 
\end{equation}
where $P(,,,)$ is the joint probability
\begin{equation}
P(w,t,w',0)=P(w,t|w',0) \cdot P(w',0).
\end{equation} 
The associated correlation function is exponential with correlation 
length $\xi
$, corresponding to a Lorentzian power spectral density (PSD)
\begin{equation}
{\rm {PSD}}(f) =  \frac{1}{f^2+\xi^2}  \quad , 
\end{equation}
where $f$ is the frequency (arbitrary units). The tails of the Lorentzian
decay, of course, as $f^{-\alpha}$ with $\alpha=2$.

In the more difficult situation when $w_0>0$, simulations were used
to solve Eq. (5). The power-law exponent 
$\alpha$ was estimated by means of a straight-line fit of
the power spectral density (represented on doubly logarithmic coordinates)
after 1024 iterations of a given value of $w_i$. 100 such runs were
carried out for each value of $w_0$. The average values (dots) and 
standard deviations (error bars) of $\alpha$ are presented in Fig. 2 
as a function of $w_0$. The exponent clearly decreases
with increasing $w_0$, revealing the onset of fractal behavior. 
It assumes a maximum value of about 
$1.75\pm0.25$ at $w_0=0$, which is about one standard deviation below 
its expected value of $2$. We expect that the discrepency results from
finite data length. 

Finally, in Fig. 3 we illustrate the simulated amplitude histograms for 
this multiplicative process ($N=2$) with the bound $w_0$ as a parameter. 
The distributions gradually move from an arcsine-like form for $w_0=0$ 
to a rather
rather peaked form for $w_0=0.775$. They are clearly non-Gaussian
for all $w_0$, leaving no doubt that the resulting process 
is not equivialent to (fractal) Gaussian noise.

\subsubsection{$N>2$, Numerical Results}
We now consider interactions involving more than two coupled elements. 
We have simulated Eq. (5) to obtain spectral densities 
for various values of $w_0$ and $N$. 
The power spectral densities for $w_0=0.6,0.7,0.8$ are shown in the
three panels of Fig. 4, for $N=2,10, 50$ elements respectively. 
On these doubly logarithmic coordinates, it
is clear that the processes all exhibit power-law spectra. 
Estimates of the fractal exponents $\alpha$ over the frequency 
range $10 < f < 512$ (arbitrary units) are provided in Table I. 
The exponents displayed in the table clearly decrease
with increasing $w_0$, revealing the onset of fractal behavior. The 
results for $N=10, 50$ do not differ substantially from those for
$N=2$ (also shown in Fig. 2). It is also apparent from Table I that
$\alpha(w_0=0)$ assumes a maximum value of about 
$1.75\pm0.25$ which, just as for $N=2$, is about one standard deviation 
below its expected value of $2$. We expect that the discrepency here too 
results from finite data length.
 
The results embodied in Fig. 4 constitute 
a key finding of our work:  coupled arrays of multiplicative random 
processes that are subject to constraints exhibit
power-law behavior in the power spectral density. 

The amplitude histograms for $w_0=0.0,0.2,0.4$ are shown in 
the two panels of Fig. 5, for $N=10, 50$ elements, respectively. On these
doubly logarithmic coordinates $P(w)$ is seen to exhibit power-law behavior,
in agreement with the asymptotic result given in Eq. (4) \cite{Levy96}. 
This is true even for $w_0=0$. In no case was power-law amplitude behavior
evident for two elements (the amplitude histograms for $N=2$ are
displayed in Fig. 3). It is therefore clear that 
the emergence of fractal amplitude behavior arises from the 
presence of a sizeable number of interacting elements. This is another
key finding of our work.
 
The behavior of $\alpha$, over a range of $w_0$ that is of interest 
for the example provided in the next section, is plotted
in Fig. 6 for several values of $N$. These curves can be well fit by 
a three-parameter function of the form
\begin{equation}
\label{fit}
\alpha(w_0) = c_1-(w_0+c_2)^{c_3} \quad ;
\end{equation}
the parameter values $c_1,c_2,c_3$ are provided in Table II. It is
clear from Fig. 6 that the exponent $\alpha(w_0)$ depends strongly
on the value of the bound $w_0$. 

A crucial observation to be gleaned from Table I and Fig. 6 is the 
decrease 
in the power-law exponent, and the concomitant departure of the
correlations from exponential form, that herald the onset of fractal
behavior as the lower bound $w_0$ increases, whatever the value
of $N$. Therein resides
the origin of the $1/{f{-\alpha}}$ behavior 
in a coupled multiplicative system with conservation constraints. 

\section{Example: Estimation of Fractal Exponents for Sequences of
Cochlear Nerve-Fiber Action Potentials}

The transmission of auditory information from the mammalian hair-cell
transducer to higher centers in the brain is mediated by
a flux of neurotransmitter molecules. These molecules are produced
in the inner hair cell at a certain limited rate. After exocytosis and 
diffusion across the synaptic cleft, they
are distributed among the roughly 10-20 primary cochlear-nerve 
fibers (CNFs) that innervate each hair cell, and result in the firing
of sequences of nerve spikes that travel on up the auditory 
pathway \cite{PIC88}.

Assuming that the model presented here is applicable for describing
this process, we associate the amplitude $w_i$ in Eq. (5)
with the neurotransmitter flux reaching one of the $N$ nerve fibers 
that synapse on a particular inner hair cell. Since the neural
firing rate is proportional to the neurotransmitter flux, we can
estimate $w_0$ by determining the lowest local firing rate
from a CNF spike train. A simple way to achieve this is
to divide the spike train into contiguous segments of $T$ sec, and
then to define $w_0$ as the ratio of the minimum firing rate 
observed over the
entire data set, to the average firing rate $\rho$. Given $w_0$
and $N$ (which determines $c_1$, $c_2$, and $c_3$ in accordance
with Table II), the expected fractal exponent $\alpha_{\rm{th}}$
is provided by Eq. (15) (see Fig. 6).

We carried out this procedure for cat CNF spike trains obtained in
nine experiments \cite{LOWE96,CAR96}, using a time window 
of $T=10$ sec and $N=10$ interacting elements. The 
results for $\alpha_{\rm{th}}$ are given in Table III, along 
with the measured fractal exponent $\alpha_{\rm{exp}}$ obtained
from the spectra and Allan factors of the spike 
trains themselves \cite{LOWE96,LOWE97}. 
The results are in substantial, and surprising, agreement indicating
that it is worthwhile to further pursue this line of reasoning.

\clearpage

{\large  TABLES}
\begin{table}[h]
\begin{center}
\begin{tabular}{ l c c c c c }
\hline
Array size    & $\alpha(w_0=0.0)$    & $\alpha(w_0=0.1)$  & $\alpha(w_0=0.2)$      & $\alpha(w_0=0.3)$     & $\alpha(w_0=0.4)$       \\ 
\hline
\hline
 $N=2$     & $1.73(25)$ &  $1.74(9)$  & $1.73(7)$  & $1.69(8)$  & $1.66(7)$  \\
\hline
 $N=10$    & $1.70(23)$ &  $1.75(12)$ & $1.69(11)$ & $1.71(9)$  & $1.70(10)$ \\
\hline
 $N=50$    & $1.72(26)$ &  $1.74(18)$ & $1.74(18)$ & $1.70(14)$ & $1.69(15)$ \\
\hline
 & & & & &  \\
\hline
Array size       & $\alpha(w_0=0.5)$     & $\alpha(w_0=0.6)$     &  $\alpha(w_0=0.7)$   & $\alpha(w_0=0.8)$ &   \\ 
\hline
\hline
 $N=2$     &  $1.64(7)$  & $1.52(7)$  & $1.27(7)$ & $0.71(6)$ &\\
\hline
 $N=10$    &  $1.63(10)$ & $1.50(9)$  & $1.20(10)$& $0.43(8)$ &\\
\hline
 $N=50$    &  $1.63(12)$ & $1.52(11)$ & $1.13(11)$& $0.13(8)$ &\\
\hline
\end{tabular}
\end{center}
\caption{Fractal exponents $\alpha$ for the multiplicative 
stochastic model with 
$N=2,10,50$ coupled variables, for various values of the lower bound 
$w_0$ 
operative on the individual variables. The numbers in parentheses
indicated standard deviations. The top-most row ($N=2$) corresponds to
the data plotted in Fig. 2. For $w_0=0$, a
nominal value $\alpha=1.72\pm0.25$ emerges independently of the
number of elements $N$, suggesting that the most salient feature of
the model responsible for fractal spectral behavior is the 
constraint.     }
\label{tab1}
\end{table}
\begin{table}[h]
\begin{center}
\begin{tabular}{ r c c r }
\hline
$N$                        & $c_1$ & $c_2$ & $c_3$    \\ 
\hline
\hline
$2$                        & $1.74$& $0.201$& $7.3$   \\
$10$                       & $1.75$& $0.235$& $8.1$   \\
$50$                       & $1.72$& $0.250$& $10.0$   \\
\hline
\end{tabular}
\end{center}
\caption{ Parameters $c_1,c_2,c_3$ that provide the best fits 
of Eq. (15) to the curves shown in Figure 6. }
\label{tab2}
\end{table}

\begin{table}[h]
\begin{center}
\begin{tabular}{ l  c c c c}
\hline
Recording   &  Mean firing rate $\rho$  &  $w_0$  & $\alpha_{{\rm theory}}$  &
 $\alpha_{{\rm experiment}}$  \\ 
\hline
\hline
  L1802   &   $71.5$ & $0.7466$ & $0.89$  &$0.90$\\
  L1805   &   $96.8$ & $0.7888$ & $0.54$  &$0.55$\\
  L1903   &   $75.1$ & $0.7722$ & $0.69$  &$0.60$\\
  L1907   &   $91.7$ & $0.7561$ & $0.82$  &$0.89$\\
  L1908   &   $106.4$& $0.7734$ & $0.68$  &$0.89$\\
  L1909   &   $98.4$&  $0.7612$ & $0.78$  &$1.20$\\
  L1910   &   $91.5$&  $0.7662$ & $0.74$  &$0.64$\\
  R0702   &   $14.1$&  $0.6980$ & $1.18$  &$1.18$\\
  R0703   &   $93.6$&  $0.6711$ & $1.30$  &$1.57$\\
\hline
\end{tabular}
\end{center}
\caption{ Comparison between the predicted fractal exponents  
$\alpha_{{\rm th}}$ obtained using the  
multiplicative stochastic model with $N=10$ coupled, constrained, and 
interchangeable variables; and the fractal exponents $\alpha_{{\rm exp}}$ 
estimated directly from the spike trains. The agreement is 
unexpectedly good.  }
\label{tab3}
\end{table}

\clearpage

{\large  FIGURE CAPTIONS}
\begin{figure}[htb]
{\baselineskip=13pt
Figure 1.~Computed form for the probability matrix $W$ when the random
variable $\lambda$ is drawn from a uniform probability distribution. 
Only values of $w$ constrained from below by $w > 0$ are permitted. 
\baselineskip=15pt}
\end{figure}
\begin{figure}[htb]
\label{fig2}
{\baselineskip=13pt
Figure 2.~Exponent of the power spectral density $\alpha$ as 
a function of the lower bound constraint on the amplitude $w_0$ 
for a constrained two-element stochastic multiplicative process. 
The decrease of the exponent from its nominal value of 2 at $(w_0=0)$
indicates a transition from Lorentzian to fractal behavior.
\baselineskip=15pt}
\end{figure}
\begin{figure}[htb]
\label{fig3}
{\baselineskip=13pt
Figure 3.~Simulated amplitude histograms $P(w)$ for various values of
$w_0$ when $N=2$. The 
curves are symmetrical about unity and are clearly non-Gaussian. 
Power-law behavior of the amplitude histogram $P(w)$ [as provided in 
Eq. (4)], emerges only as $N$ increases, as will become clear from Fig. 5.
\baselineskip=15pt}
\end{figure}
\begin{figure}[htb]
\label{fig4}
{\baselineskip=13pt
Figure 4.~Power spectral densities (PSDs) with
$w_0$ as a parameter for three values of $N$: $N=2,10, 50$ (upper, 
middle, and lower panels respectively). Power-law behavior is
observed over the frequency range of $10 < f < 512$ (arbitrary units),
even in the case when there are only two elements. 
\baselineskip=15pt}
\end{figure}
\begin{figure}[htb]
\label{fig5}
{\baselineskip=13pt
Figure 5.~Amplitude histograms with
$w_0$ as a parameter for two values of $N$: $N=10, 50$ shown in 
different panels. The presence of power-law behavior in this figure
contrasts with its absence when there are only two interacting
elements (see Fig. 3).
\baselineskip=15pt}
\end{figure}
\begin{figure}[htb]
\label{fig6}
{\baselineskip=13pt
Figure 6.~$\alpha$ as a function of $w_0$, over a limited range,
for several values of $N$.
$\alpha$ was estimated by means of a straight-line fit of
the power spectral density (represented on doubly logarithmic coordinates)
after 1024 iterations of a given value of $w_i$. 100 such runs were
carried out for each value of $w_0$. Average values are shown as dots 
and standard deviations as error bars. The curve for $N=2$ is a portion
of the one plotted in Fig. 2. The lines are to guide the eye.
\baselineskip=15pt}
\end{figure}

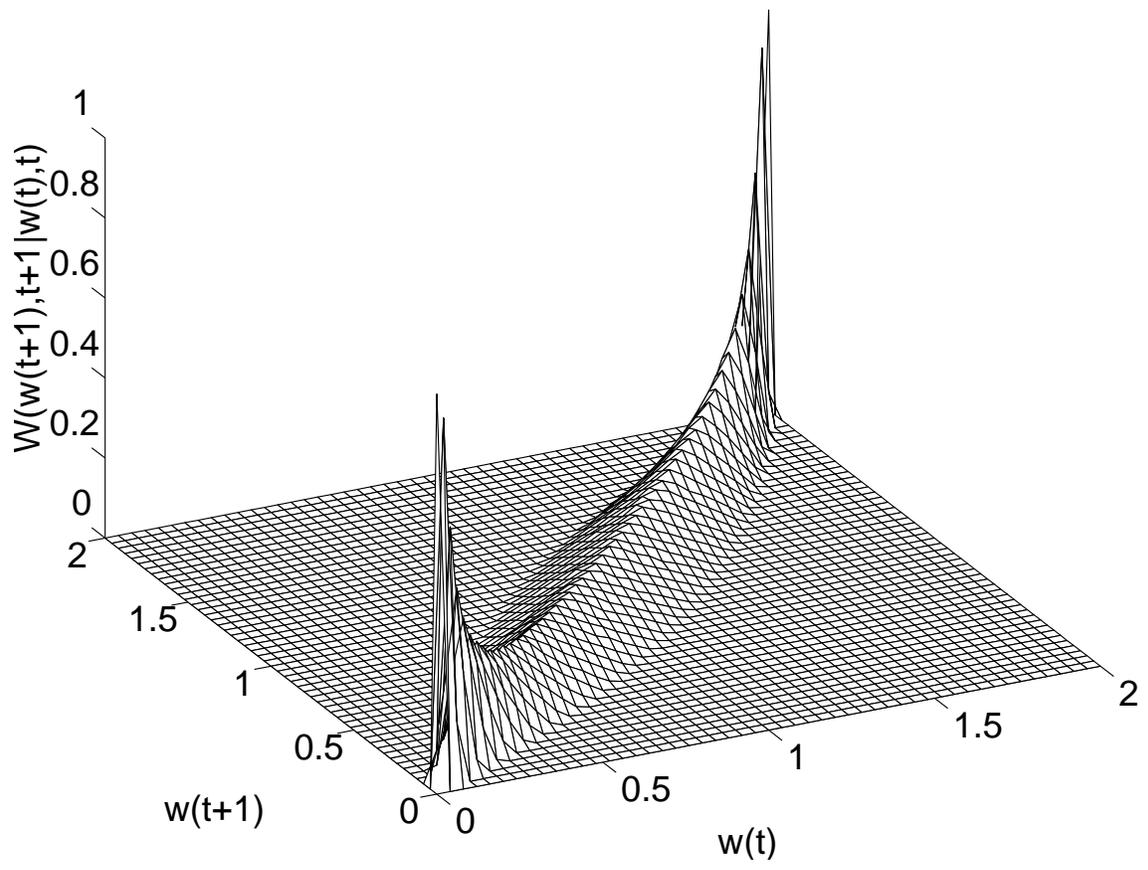
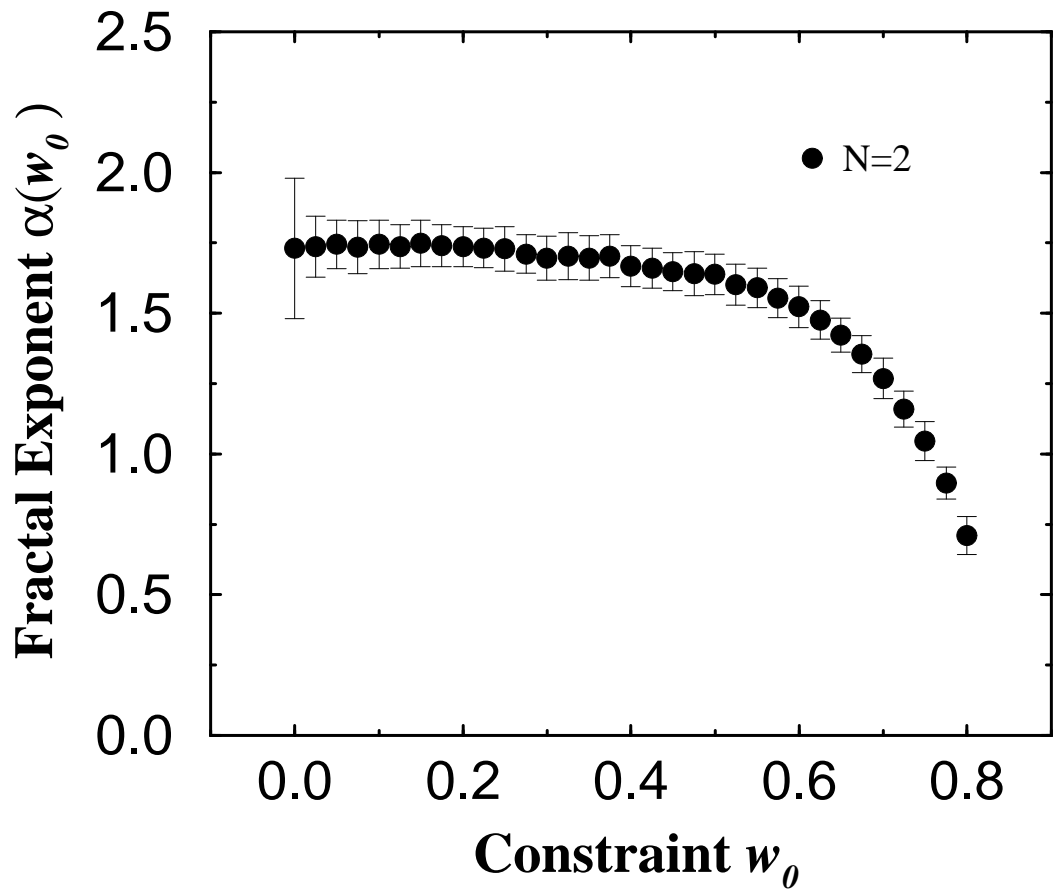
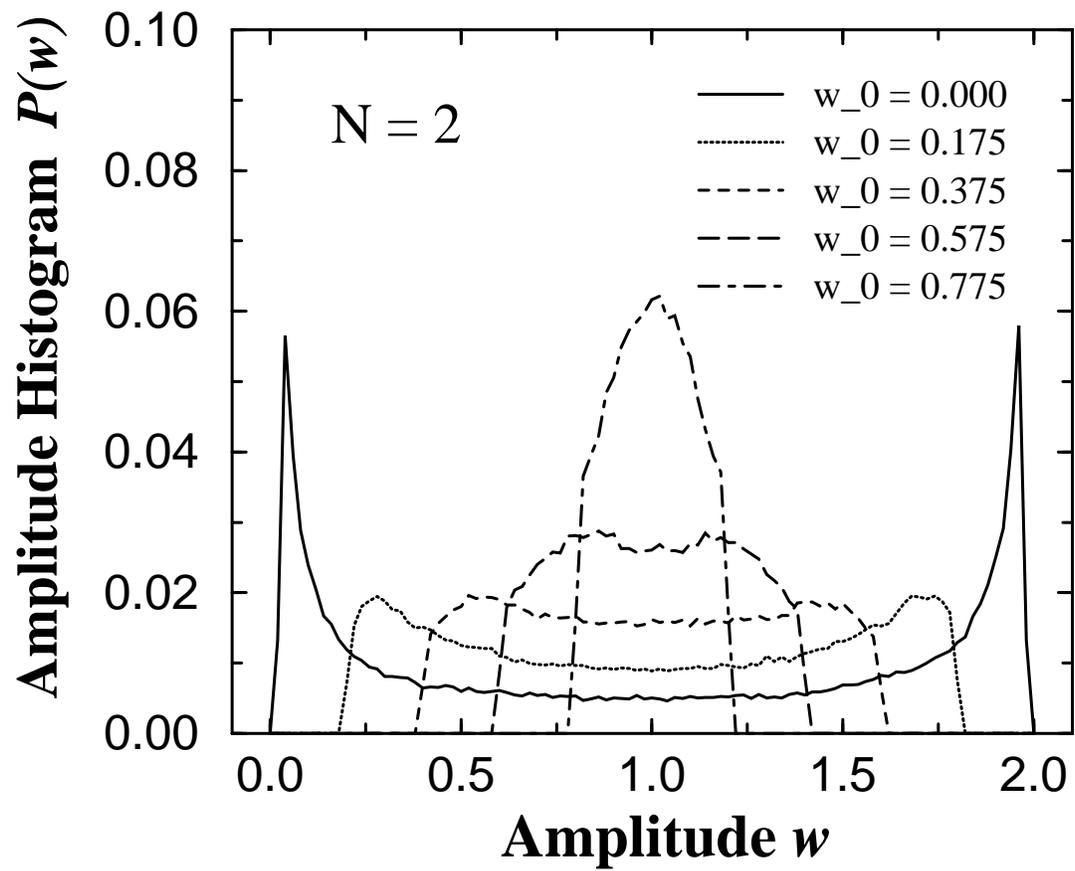
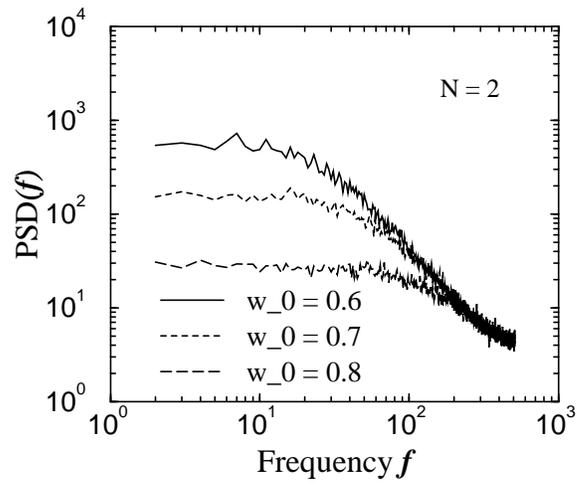
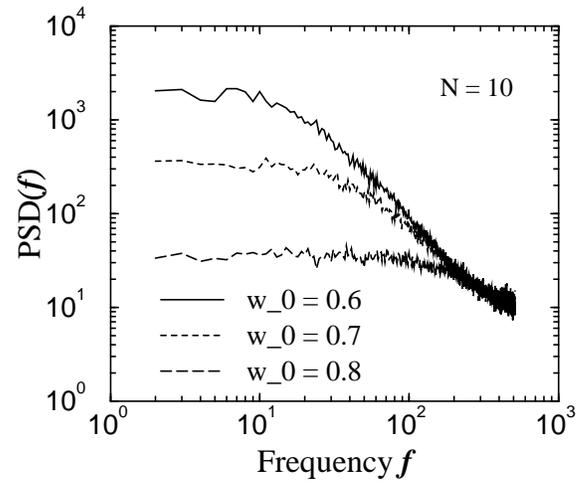
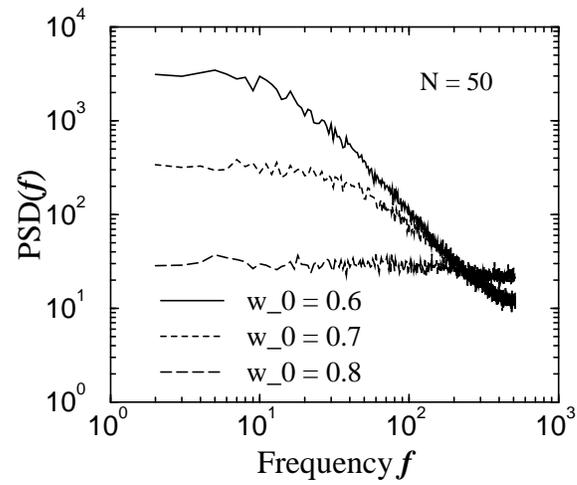
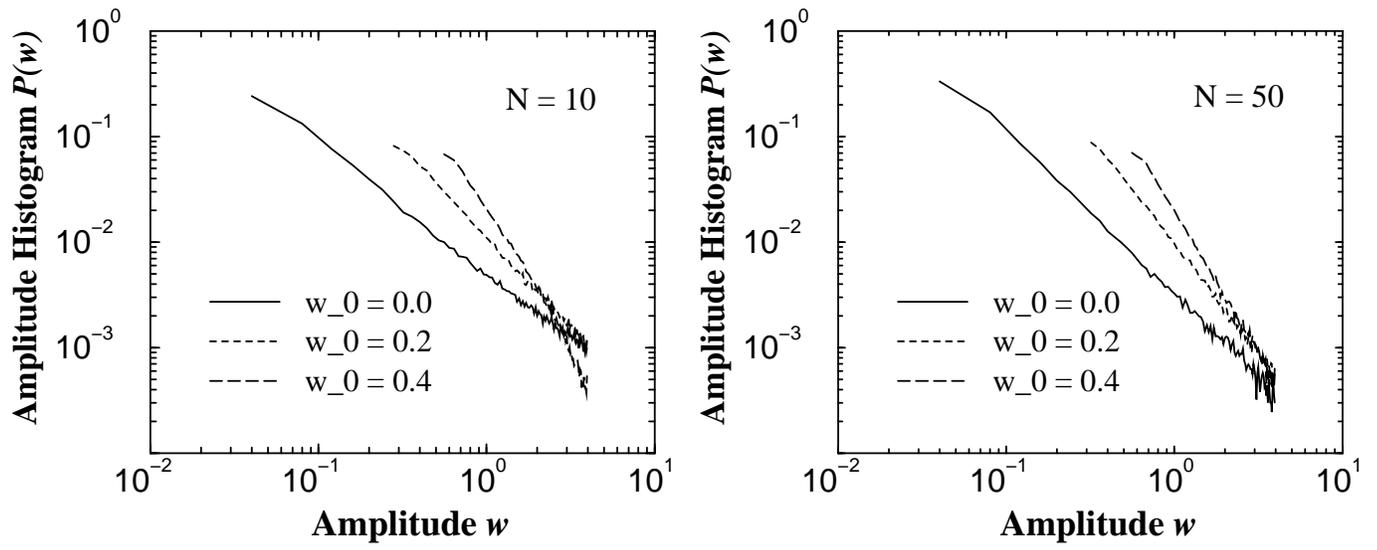
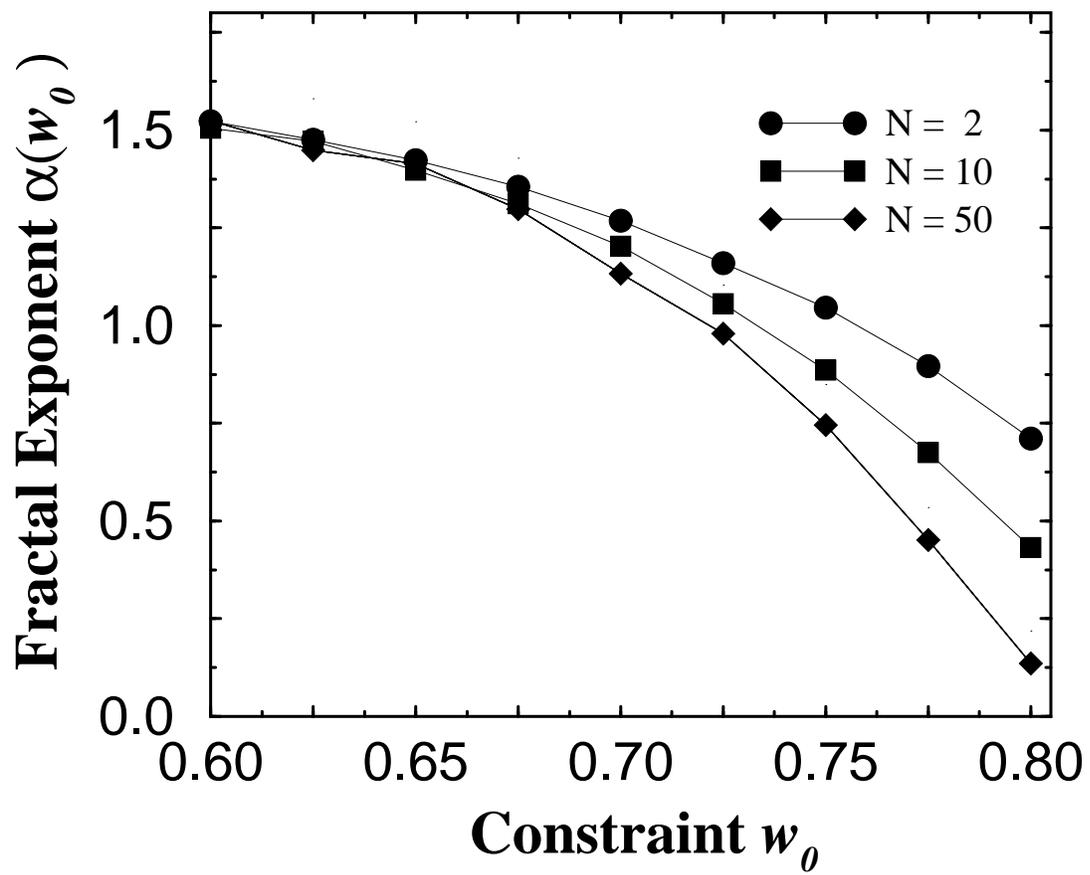
\end{document}